\documentclass[%
 reprint,
 amsmath,amssymb,
 aps,
]{revtex4-1}

\usepackage{graphicx}% Include figure files
\usepackage{dcolumn}% Align table columns on decimal point
\usepackage{bm}% bold math
\begin{document}

\preprint{APS/123-QED}

\title{Anisotropic Effective Mass}% Force line breaks with \\
\thanks{Thanks to: Dr. Leon Altschul and Prof. Shlomo Ruschin for helpfull discussions}%
\author{Viktor Ariel }
\affiliation{%
 Department of Physical Electronics\\
 Tel-Aviv University
}%

\begin{abstract}
In this work, we derive a three-dimensional effective mass that is suitable for treatment of electrons in anisotropic semiconductors.
We show that it is possible to define a scalar anisotropic three-dimensional mass that reduces to a one-dimensional definition for isotropic materials.  We extend the effective mass definition and apply it to electrons in non-parabolic semiconductors and demonstrate that the effective mass is generally dependent on particle energy.  Finally, we show  that the proposed definition of the effective mass is compatible with the experimentally measured cyclotron mass in isotropic materials while leading to a more accurate result in anisotropic semiconductors.\\

\end{abstract}

\pacs{Valid PACS appear here}% PACS, the Physics and Astronomy
\maketitle

%\tableofcontents

\section{\label{sec:level1}Introduction\protect\\}

Recently,   we presented  a one-dimensional theoretical definition of the effective mass \cite{ArielArxiv1} and   \cite{ArielArxiv2}.  We applied this definition to isotropic three-dimensional semiconductor materials and free relativistic particles \cite{ArielArxiv3}. In this paper,  we would like to propose a three-dimensional definition of the effective mass that can be applied to electrons in both isotropic and anisotropic materials. We extend the definition of the effective mass to non-parabolic semiconductor materials where the dispersion relation is given by the simplified Kane model \cite{Seeger},  \cite{Zawadzki}. Finally, we  show that the cyclotron  mass  measured in semiconductors \cite{Seeger}, \cite{Shockley}  can be considered an approximation of the anisotropic mass defined here.

\section{Definition of Anisotropic Mass } % mass definition

Previously, we obtained a one-dimensional definition of the effective mass by a semi-classical association  of  particles with  wave-packets. 

We identify  the particle momentum  with the crystal momentum $p=\hbar k$ and  associate particle velocity $v$ with the group velocity of the wave-packet. 

\begin{equation}
 v\simeq  \frac {\partial E} { \partial  p} \ .
\label{eq:group}
\end{equation}

The effective mass appears as a proportionality factor between the particle momentum and the group velocity of the wave-packet
\begin{equation}
 p \simeq  m v \ .
\label{eq:momentum}
\end{equation}

Based on equations  (\ref{eq:group})  and   (\ref{eq:momentum})   we  define the effective mass in one dimension as
\begin{equation}
m = \frac {p} {v} = \frac {p} {\partial E/ \partial p} .
\label{eq:mass-1d}
\end{equation}

The most general three-dimensional relationship between particle momentum and velocity based on (\ref{eq:momentum}) can be written in the form:

\begin{equation}
  p_i \simeq  m_{ij} v_j \ ,
\label{eq:mass-3d}
\end{equation}
where $m_{ij}$ is the effective mass tensor and $p_i$ and $v_j$ are components of the three-dimensional momentum, $\boldsymbol p$, and group velocity, $\boldsymbol v$, vectors. Here the group velocity can be defined as: 
\begin{equation}
\boldsymbol v = \boldsymbol{\nabla}_p E\ .
\label{eq:group-gradient}
\end{equation}
Any three-dimensional definition of mass should converge to the one-dimensional definition (\ref{eq:mass-1d}) for one-dimensional particle movement and for three-dimensional isotropic materials. However, 
there appears an ambiguity in the general equation (\ref{eq:mass-3d}) due to the fact that different anisotropic  definitions of $m_{ij}$ may lead to the same isotropic relation (\ref{eq:mass-1d}).  

Also, we seek guidance from experimental measurements of the effective mass based on cyclotron resonance  since it is one of the most common methods of measuring carrier effective mass in semiconductors.
We note that in  cyclotron resonance measurements of both isotropic and  anisotropic materials, the measured carrier mass appears  as a scalar quantity \cite{Seeger}. Also, both mass and mass density are scalar quantities in classical mechanics.

We  therefore look for a scalar definition of the effective mass.  It seems possible to use the ratio of the absolute values of the momentum and group velocity as a scalar three-dimensional definition of the effective mass:

\begin{equation}
 m =\frac {p} {v}=\frac {|{\boldsymbol p}|} {|{\boldsymbol v}|}  = \frac {|{\boldsymbol p}|} { |  \boldsymbol{\nabla}_p E |}\  .
\label{eq:mass-scalar}
\end{equation}

Note that the above three-dimensional expression (\ref{eq:mass-scalar}) clearly reduces to the regular one-dimensional definition (\ref{eq:mass-1d}) for isotropic materials. We anticipate that in anisotropic materials the effective mass will be a function of the direction of particle movement relative to the main crystal axes. Since the particle momentum and group velocity are not generally collinear in anisotropic materials, we expect that the effective mass will be a function of the angle between particle velocity and momentum.

\section{Parabolic Anisotropic mass } % parabolic mass

We would like to apply  the effective mass definition (\ref{eq:mass-scalar})  to particles with a parabolic anisotropic dispersion relation.  Let us consider the electron kinetic energy in anisotropic semiconductors, such as Si and Ge, which can be expressed \cite{Seeger} as:

\begin{equation}
E = \frac {p_x^2} {2 m_x} + \frac {p_y^2} {2 m_y} + \frac {p_z^2} {2 m_z}\ ,
\label{eq:energy-anisotropic}
\end{equation}
where $p_x$, $p_y$, and $p_z$ are projections of the electron momentum on the main crystal axes, and $m_x$, $m_y$, and $m_z$ are positive constant coefficients corresponding to the electron rest masses along the main crystal axes.

First, we calculate components of the group velocity by differentiation of (\ref{eq:energy-anisotropic}) and using (\ref{eq:group-gradient}):
\begin{equation}
v_x =  \frac {p_x} {m_x} ,  \ v_y=  \frac {p_y} {m_y} , \ v_z=  \frac {p_z} {m_z} \ .
\label{eq:group-components}
\end{equation}
We can write  equivalently for the components of the particle momentum:
\begin{equation}
p_x =  {m_x} {v_x} ,  \ p_y=   {m_y} {v_y} , \ p_z=  {m_z} {v_z} \ .
\label{eq:momentum-components}
\end{equation}

Let us define directional cosines between particle momentum and the main crystal axes as $\alpha_p$, $\beta_p$, and $\gamma_p$, then 
\begin{equation}
p_x = \alpha_p {  p} , \ p_y = \beta_p { p} , \  p_z = \gamma_p {  p}\ .
\label{eq:momentum-proj}
\end{equation}
Similarly, we can define directional cosines for the particle velocity:
\begin{equation}
v_x = \alpha_v  v , \ v_y = \beta_v   v , \  v_z = \gamma_v v \ ,
\label{eq:velocity-proj}
\end{equation}
with the following auxiliary condition:
\begin{equation}
\alpha_p^2 + \beta_p^2  +\gamma_p^2 =   \alpha_v^2 + \beta_v^2 + \gamma_v^2 =  1 \ .
\label{eq:cos-square}
\end{equation}
This leads to the absolute value of the group velocity: 
\begin{equation}
  v =  \sqrt {\frac {p_x^2} {m_x^2}  + \frac {p_y^2} {m_y^2} + \frac {p_z^2} {m_z^2} }
 = p\  \sqrt { \frac  {\alpha_p^2} {m_x^2}  + \frac  { \beta_p^2} {m_y^2}  + \frac  {\gamma_p^2} {m_z^2}  } \ .
\label{eq:group-abs}
\end{equation}

Similarly for the absolute value of the momentum:
\begin{eqnarray}
p=  \sqrt {  m_x^2 v_x^2 +  m_y^2 v_y^2 +  m_z^2 v_z^2} \nonumber\\
= v \sqrt {  m_x^2 \alpha_v^2 +  m_y^2 \beta_v^2 +  m_z^2 \gamma_v^2}\ .
\label{eq:momentum-abs}
\end{eqnarray}
Then from (\ref{eq:group-abs}) and (\ref{eq:momentum-abs}) we obtain the parabolic effective mass, $m_0$, as
\begin{eqnarray}
m_0  =\frac { p} {  v}=  
\left [{ \frac  {\alpha_p^2} {m_x^2}  + \frac  { \beta_p^2} {m_y^2}  + \frac  {\gamma_p^2} {m_z^2}  }\right ] ^{-1/2} \nonumber \\
=\sqrt {  m_x^2 \alpha_v^2 +  m_y^2 \beta_v^2 +  m_z^2 \gamma_v^2}\ .
\label{eq:mass-parab}
\end{eqnarray}

As expected from the parabolic dispersion relation (\ref{eq:energy-anisotropic}), the effective mass (\ref{eq:mass-parab}) is not a function of energy. However, it depends on the angle between the particle momentum and crystal orientation as well as the constant rest mass coefficients along the main crystal axes. In isotropic materials, $m_x=m_y=m_z = m_0$, and  (\ref{eq:mass-parab}) reduces to the one-dimensional isotropic model (\ref{eq:mass-1d}) as expected.
When a particle is moving along one of the crystal axes, for example $p=p_x, \ p_y=0, \ p_z=0$, we obtain from (\ref{eq:mass-parab}) that $m_0=m_x$ as expected. Thus, the model reduces to a correct solution for one-dimensional motion and for particles in  isotropic materials (\ref{eq:mass-1d}).

Note that based on  (\ref{eq:momentum-proj}),  (\ref{eq:velocity-proj}),  and   (\ref{eq:mass-parab}), the following relations hold for the directional cosines:
\begin{equation}
\alpha_v = \alpha_p\frac {m_0}  {m_x} , \ \beta_v = \beta_p\frac {m_0}  {m_y} , \  \gamma_v = \gamma_p\frac {m_0}  {m_z}  \ .
\label{eq:cos-vp}
\end{equation}

If we define the angle between particle velocity and momentum, $\theta$, we can calculate it from  (\ref{eq:momentum-proj}),  (\ref{eq:velocity-proj}),   and (\ref{eq:cos-vp}): 

\begin{eqnarray}
\cos(\theta) =\frac { \boldsymbol p\cdot \boldsymbol v} { p\  v} =  
\alpha_p\alpha_v+\beta_p\beta_v+\gamma_p\gamma_v  \nonumber \\
=\frac  {m_0} {m_x} \alpha_p^2  + \frac  {m_0} {m_y} \beta_p^2   + \frac  {m_0} {m_z} \gamma_p^2   \nonumber \\
=\frac  {m_x} {m_0} \alpha_v^2  + \frac  {m_y} {m_0} \beta_v^2   + \frac  {m_z} {m_0} \gamma_v^2\ ,
\label{eq:cos}
\end{eqnarray}
where based on (\ref{eq:cos-square}) $\cos(\theta) $ is always positive.
We can see from (\ref{eq:cos}) that the effective mass (\ref{eq:mass-parab})  helps to define the angle between particle momentum and velocity in anisotropic materials. Alternatively from (\ref{eq:cos}), we can show that the effective mass depends on the angle between particle velocity and momentum:
\begin{eqnarray}
 m_0 = {\cos(\theta)} \  \left( {\frac  { \alpha_p^2} {m_x}  + \frac  {\beta_p^2 } {m_y}   + \frac  {\gamma_p^2} {m_z} } \right) ^{-1} \nonumber \\
= \frac {\left(  {m_x} { \alpha_v^2}  +  {m_y}  {\beta_v^2 }   +   {m_z}  {\gamma_v^2} \right)}   {\cos(\theta)} \ .
\label{eq:mass-cos}
\end{eqnarray}

Therefore, we demonstrated that the electron effective mass defined by (\ref{eq:mass-scalar}), (\ref{eq:mass-parab}), and  (\ref{eq:mass-cos}) depends on the direction of the particle movement relative to the crystal orientation as well as the angle between particle velocity and momentum.

\section{Non-Parabolic Anisotropic mass } % non-parabolic mass

Next, we would like to apply the effective mass definition (\ref{eq:mass-3d}) to electrons in a non-parabolic anisotropic semiconductor. The non-parabolicity is typically observed for the electron kinetic energy comparable to the semiconductor bandgap, $E_G$.  We use the non-parabolic dispersion relation based on the simplified two-band Kane model  \cite{Seeger}, \cite{Zawadzki}:

\begin{equation}
E+\frac {E^2} {E_G} = \frac {p_x^2} {2 m_x} + \frac {p_y^2} {2 m_y} + \frac {p_z^2} {2 m_z}\ ,
\label{eq:energy-nonp}
\end{equation}
We can solve for each component of the group velocity by differentiating both sides of (\ref{eq:energy-nonp}), resulting for example for the x-component:
\begin{equation}
v_x = \frac {\partial E} { \partial p_x} =  \frac {p_x} {m_x} \left( 1 + \frac {2E }{ E_G} \right)^{-1}\ .
\label{eq:group-x}
\end{equation}
Then similar to  (\ref{eq:group-abs}), we derive for the absolute value of the velocity:

\begin{equation}
v = p\ \left( 1 + \frac {2E }{ E_G} \right)^{-1} \sqrt { \frac  {\alpha_p^2} {m_x^2}  + \frac  { \beta_p^2} {m_y^2}  + \frac  {\gamma_p^2} {m_z^2}  } \ .
\label{eq:group-nonp}
\end{equation}
Finally, we obtain the non-parabolic anisotropic effective mass:

\begin{equation}
m =  \left( 1 + \frac {2E }{ E_G} \right) m_0 \ ,
\label{eq:mass-nonp}
\end{equation}
where $m_0$ is the parabolic mass given by  (\ref{eq:mass-parab}) or (\ref{eq:mass-cos}).

We can obtain a relation for the directional cosines in non-parabolic materials, for example, for the $x$-direction using  (\ref{eq:velocity-proj}),  (\ref{eq:group-abs}), and (\ref{eq:group-x}):
\begin{equation}
\alpha_v =\alpha_p\ \left( 1 + \frac {2E }{ E_G} \right)^{-1} \frac {m} {m_x} = \alpha_p\frac {m_0} {m_x} \ .
\label{eq:group-x-nonp}
\end{equation}
Since this is the same result as for parabolic materials (\ref{eq:cos-vp}), the angle between velocity and momentum is the same as in parabolic materials and given by (\ref{eq:cos}).

Expressing the effective mass as a function of the angle between velocity and momentum in non-parabolic materials, we obtain from (\ref{eq:mass-nonp}) and (\ref{eq:group-x-nonp}):
\begin{eqnarray}
 m = {\cos(\theta)} \left( 1 + \frac {2E }{ E_G} \right)\  \left( {\frac  { \alpha_p^2} {m_x}  + \frac  {\beta_p^2 } {m_y}   + \frac  {\gamma_p^2} {m_z} } \right) ^{-1} \nonumber \\
= \left( 1 + \frac {2E }{ E_G} \right)\frac {\left(  {m_x} { \alpha_v^2}  +  {m_y}  {\beta_v^2 }   +   {m_z}  {\gamma_v^2} \right)}   {\cos(\theta)} \ .
\label{eq:mass-cos-nonp}
\end{eqnarray}
As we can see from  (\ref{eq:mass-nonp}) and (\ref{eq:mass-cos-nonp}) in non-parabolic anisotropic materials, the electron effective mass is a  function of the crystal orientation and the angle between particle velocity and momentum, similar to electrons in parabolic materials. The non-parabolic effective mass is a function of particle energy unlike the effective mass in parabolic semiconductors.

\section{Comparison with cyclotron mass } % Cyclotron Mass

Previously we demonstrated  \cite{ArielArxiv2}   that in isotropic materials, the electron effective mass defined by (\ref{eq:mass-1d}) corresponds to the measured cyclotron effective mass. The situation is more complex in anisotropic semiconductors. Let us consider an electron  under the influence of the constant magnetic field  ${ \boldsymbol  B}$. Writing  Lorentz force   we obtain:

\begin{equation}
\frac {d{\boldsymbol p} } {d t}=e\  ({\boldsymbol v}\times { \boldsymbol  B}) .
\label{eq:lorentz}
\end{equation}
Assume that the electron  follows a closed orbit in the constant magnetic field. Then, integrated over one period of particle motion and  using the procedure similar to  \cite{Seeger} and  \cite{Shockley} we can write:
\begin{equation}
 p = e B \oint  {v_\perp} dt  ,
\label{eq:momentum-lorentz}
\end{equation}
where $v_\perp$  is the component of the electron velocity perpendicular to the magnetic field. In general, we can define the cyclotron mass based on (\ref{eq:mass-scalar}) and (\ref{eq:momentum-lorentz}) as:

\begin{equation}
m =\frac {p} {v} =  {e B}  \ \frac {\oint  {v_\perp} dt}  {v} \ .
\label{eq:mass-lorentz}
\end{equation}

For isotropic materials the particle is moving in a circle and $|v_\perp| = v $ is a constant.  Consequently, (\ref{eq:momentum-lorentz}) is easily integrated leading to the cyclotron effective mass:
\begin{equation}
 m_c \simeq \frac {e B} {\omega_c},
\label{eq:mass-cyclotron}
\end{equation}
where $\omega_c$ is the cyclotron frequency.

In anisotropic materials, the electron trajectory is not cirular and  $v$ and  $m_0$ are not constant during a period of particle rotation, as can be seen from  (\ref{eq:group-abs}) and (\ref{eq:mass-parab}) . Thus, the cyclotron mass (\ref{eq:mass-cyclotron}) represents an approximate average value of the effective mass defined by  (\ref{eq:mass-scalar})  and (\ref{eq:mass-lorentz})  over one period of particle rotation.

If we  define the directional cosines between the magnetic field ${ \boldsymbol  B}$ and the crystal axes as $\alpha_b$, $\beta_b$, and $\gamma_b$, then based on (\ref{eq:mass-scalar}) and using a derivation from \cite{Seeger} we obtain for parabolic materials: 

\begin{equation}
m_c = \bar {m}_0 = \left [\frac {m_x m_y m_z} {\alpha_b^2 m_x+\beta_b^2 m_y+\gamma_b^2 m_z}   \right ] ^{1/2}\ ,
\label{eq:mass-projb}
\end{equation}
where $\bar {m}_0$ is the average value of the electron effective mass over one period of cyclotron motion.
The above expression was experimentally confirmed by cyclotron resonance measurements in several anisotropic materials \cite{Seeger}.

Let us assume that a constant magnetic field is applied in $z$-direction and that that the electron is moving in the $xy$-plane. Then, from (\ref{eq:mass-parab}) we obtain the value of the parabolic effective  mass:

\begin{equation}
m_0=\sqrt {  m_x^2 \alpha_v^2 +  m_y^2 \beta_v^2}\ .
\label{eq:mass-exact}
\end{equation}
Note from  (\ref{eq:mass-exact}) that during particle rotation in the $xy$-plane the effective mass changes between $m_x$ and $m_y$.

The cyclotron mass given by (\ref{eq:mass-cyclotron}) and (\ref{eq:mass-projb}) leads to a constant effective mass:
\begin{equation}
m_c = \sqrt{m_x m_y }  \ .
\label{eq:mass-approx}
\end{equation}
For isotropic materials with $m_x=m_y=m_0$, both expressions  (\ref{eq:mass-exact}) and  (\ref{eq:mass-approx}) lead to the same result $m_c=m_0$. In anisotropic materials, 
the result of (\ref{eq:mass-approx}) is clearly not accurate since it neglects fluctuations of the effective mass during a period of particle rotation while (\ref{eq:mass-exact}) produces an accurate result. Thus, the cyclotron mass  (\ref{eq:mass-approx})   can be considered an approximation of (\ref{eq:mass-exact}) averaged over one period of particle rotation.

The linear dependence of the effective mass on energy leads to electron velocity saturation at high particle energy \cite{ArielArxiv3}, which was demonstrated experimentally by cyclotron resonance measurements in non-parabolic  InSb  \cite{Zawadzki}.

Therefore, we demonstrated that the cyclotron mass defined by  (\ref{eq:mass-cyclotron}) leads to a correct result in isotropic materials. However in anisotropic materials, the cyclotron mass (\ref{eq:mass-cyclotron}) neglects mass fluctuations  over one period of particle rotation and consequently can be considered an approximation of the accurate anisotropic mass defined by (\ref{eq:mass-scalar}). Also, the non-parabolic effective mass depends on particle energy as was demonstrated experimentally.

\section{Conclusions}

In this work, we proposed a three-dimensional anisotropic definition of the particle effective mass. In our definition, the effective mass appears as a scalar quantity  dependent on the direction of  particle motion relative to crystal orientation and on the angle between particle velocity and momentum. We applied the anisotropic mass to electrons in non-parabolic  materials and demonstrated that the effective mass is generally a function of electron energy. We demonstrated that the new definition  reduces to the one-dimensional mass for electrons in isotropic materials.  We showed that the anisotropic mass  is compatible with the cyclotron mass in isotropic materials, while in anisotropic materials the cyclotron mass can be considered an average approximation of the more accurate anisotropic effective mass presented here.


\begin{thebibliography}{9}



\bibitem{ArielArxiv1}
V. Ariel, arXiv:1205.3995v1 [physics.gen-ph], (2012).

\bibitem{ArielArxiv2}
V. Ariel, A. Natan,  	arXiv:1206.6100v1 [physics.gen-ph], (2012).

\bibitem{ArielArxiv3}
V. Ariel, arXiv:1207.4282v1 [physics.gen-ph], (2012).



\bibitem{Seeger}
K. Seeger, 
\emph{Semiconductor Physics}
(Springer-Verlag, 1985).



\bibitem{Zawadzki}
W. Zawadzki, S. Klahn, U. Merkt, Phys. Rev. Lett.  {\bf 55}, 983 (1985).


\bibitem{Shockley}
W. Shockley, Phys. Rev., {\bf 79}, 191 (1950).




\end{thebibliography}
\end{document}